\documentclass[conference]{IEEEtran}

\usepackage[letterpaper, left=0.625in, right=0.625in, bottom=1.1in, top=0.75in]{geometry}
\makeatletter
\newcommand{\ssymbol}[1]{^{\@fnsymbol{#1}}}
\makeatother

\usepackage[caption=false]{subfig}
\ifCLASSINFOpdf
  \usepackage[pdftex]{graphicx}
\else
\fi
\usepackage{amsmath,amssymb,amsfonts}
\usepackage{amsmath}
\DeclareMathOperator*{\argmax}{argmax} 
\usepackage{textcomp}
\usepackage{algorithm}
\usepackage[noend]{algpseudocode}

\makeatletter
\def\BState{\State\hskip-\ALG@thistlm}
\makeatother

\algnewcommand\algorithmicswitch{\textbf{switch}}
\algnewcommand\algorithmiccase{\textbf{case}}
\algnewcommand\algorithmicassert{\texttt{assert}}
\algnewcommand\Assert[1]{\State \algorithmicassert(#1)}%
\algdef{SE}[SWITCH]{Switch}{EndSwitch}[1]{\algorithmicswitch\ #1\ \algorithmicdo}{\algorithmicend\ \algorithmicswitch}%
\algdef{SE}[CASE]{Case}{EndCase}[1]{\algorithmiccase\ #1}{\algorithmicend\ \algorithmiccase}%
\algtext*{EndSwitch}%
\algtext*{EndCase}%

\usepackage{enumitem}
\usepackage{amsmath}

\usepackage{float}
\begin{document}
%
\title{Energy-aware optimization of UAV base stations placement via decentralized multi-agent Q-learning}

\author{\IEEEauthorblockN{Babatunji Omoniwa, Boris Galkin, Ivana Dusparic}
\IEEEauthorblockA{CONNECT Centre, School of Computer Science and Statistics,\\
Trinity College Dublin, Ireland.\\
Emails: omoniwab@tcd.ie, galkinb@tcd.ie, ivana.dusparic@scss.tcd.ie}
}


%


\maketitle

\begin{abstract}
Unmanned aerial vehicles serving as aerial base stations (UAV-BSs) can be deployed to provide wireless connectivity to ground devices in events of increased network demand, points-of-failure in existing infrastructure, or disasters.
However, it is challenging to conserve the energy of UAVs during prolonged coverage tasks, considering their limited on-board battery capacity. Reinforcement learning-based (RL) approaches have been previously used to improve energy utilization of multiple UAVs, however, a central cloud controller is assumed to have complete knowledge of the end-devices' locations, i.e., the controller periodically scans and sends updates for UAV decision-making. This assumption is impractical in dynamic network environments with UAVs serving mobile ground devices. To address this problem, we propose a decentralized Q-learning approach, where each UAV-BS is equipped with an autonomous agent that maximizes the connectivity of mobile ground devices while improving its energy utilization. Experimental results show that the proposed design significantly outperforms the centralized approaches in jointly maximizing the number of connected ground devices and the energy utilization of the UAV-BSs.

\end{abstract}

\begin{IEEEkeywords}
Reinforcement learning, UAV base stations, energy management, wireless connectivity.
\end{IEEEkeywords}

%

\IEEEpeerreviewmaketitle

\section{Introduction}
The use of unmanned aerial vehicles (UAVs) to provide wireless connectivity to ground end-devices has received significant research attention \cite{Mozaffari2017UAV},~\cite{Zeng2019UAV}. In particular, the adjustable altitude and mobility of UAVs make them suitable candidates for flexible deployment as aerial base stations (UAV-BSs) in events of increased network demand, points-of-failure in existing infrastructure, or disasters. UAVs can adjust their position to establish reliable connectivity with ground end-devices~\cite{Mozaffari2017UAV},~\cite{Galkin2017UAVCoverageGlobecom}. 
Ground devices requiring connectivity may either be static, e.g. flood alert sensors, smart electric meters, air quality sensors, etc. or mobile, e.g. smartphones interconnected to smart earbuds, trackers and internal heart monitoring sensors in humans, parcel and vehicle tracking sensors, etc. Nevertheless, for UAVs to be effectively deployed to serve as UAV-BSs, several technical challenges must be addressed such as optimal placement \cite{Bor-Yaliniz2016UAV}, trajectory-planning~\cite{Wu2018UAVtrajectoryTransaction} and energy management~\cite{Mozaffari2017UAV, Zeng2019UAV, Yang2020UAVenergyharvesting, Liu2020UAVdistributed} of the UAVs. In particular, UAVs have limited on-board battery capacity, making it a challenge to serve as aerial base stations for extended periods of time \cite{Mozaffari2017UAV} -- \cite{Yao2019UAV}.

In~\cite{Zeng2019UAV} and \cite{Yao2019UAV}, a single UAV was deployed to support downlink wireless
communications to ground devices. Although these works show significant improvement in energy utilization, they focused on improving the energy utilization of a UAV providing coverage to static ground devices. In~\cite{Mozaffari2017UAV}, multiple UAVs were deployed to support wireless coverage in geographically large areas. This work minimized the UAVs' energy consumption, however, the  UAVs relied on periodic updates from a central cloud controller for decision making and control, making the approach unsuitable in disaster scenarios. A branch  of  machine learning  called  reinforcement  learning  (RL) was used in \cite{Liu2020UAVdistributed} and \cite{Liu2018UAV} to minimize energy consumption in UAVs. These works considered the coverage of only static ground devices. In~\cite{Liu2019UAVqlearning}, mobile ground devices were considered, however, an ideal assumption was made of the spatial knowledge of the locations of all ground devices in the network. Moreover, the works \cite{Liu2020UAVdistributed, Liu2019UAVqlearning, Liu2018UAV} ignored the impact of interference from neighbouring UAV cells. In a dynamic network environment, existing model assumptions are impractical since UAVs operate in the sky and have strong line-of-sight (LOS) channels even to distant users~\cite{Galkin2016UAV, Galkin2019UAV_TVT}, which is why UAV networks will be interference-limited rather than noise limited as in~\cite{Liu2019UAVqlearning} or interference-free as in~\cite{Liu2020UAVdistributed} and \cite{Liu2018UAV}. Motivated by the research gaps above, We consider the deployment of multiple UAVs to provide wireless connectivity to both static and mobile ground devices while taking into account interference from neighbouring UAV cells. Our two main contributions are as follows:
\begin{itemize}[leftmargin=*]
  \item We propose a decentralized Q-learning with local sensory information (DQLSI) algorithm  that relies on local observations from the UAV and the broadcast connectivity observation from neighbouring UAVs for decision making.  Here, each UAV is equipped with an autonomous agent that learns to jointly maximize the number of connected ground devices and energy utilization by UAVs while optimizing the 3D flight trajectory of each UAV over a series of time-steps.
  \item We compare our proposed DQLSI approach to the state-of-the-art centralized approaches that assume global spatial knowledge of all ground devices in a quickly-evolving network. We also evaluated the approach using two different models of ground device mobility: random walk (RW) and random waypoint (RWP)~\cite{Camp_RW_2002}. Results show that our proposed approach is robust in simultaneously providing improved energy utilization while maximizing the number of connected ground devices.
\end{itemize}
The remainder of this work is organized as follows.~In Section II, we present related work. The environment model is provided in Section III. We present DQLSI approach in section IV. In section V, we present the simulation setup, results and discussions. Section VI concludes the paper and outlines future directions.

\vspace{-3mm}
\section{Related Work}
In recent work~\cite{Zeng2019UAV}, \cite{Yao2019UAV}, a single UAV deployed as an aerial base station to provide coverage to static ground devices was shown to have a significant improvement in energy utilization. However, these works were limited by a single energy management policy, a static and an ideal coverage space. In~\cite{Mozaffari2017UAV}, an iterative algorithm was proposed to jointly optimize the three-dimensional (3D) placement and mobility of the UAV-BS, to minimize the total energy consumed by multiple UAV-BSs in the network. The work was able to improve energy utilization, however, the periodicity of control and data updates with the central control station may result in increased energy consumption for the UAVs. It is worthwhile to look into alternatives of relying on a central controller in the advent of a disaster scenario where the ground infrastructure may be damaged~\cite{Liu2020UAVdistributed}. In particular, to reduce cost and complexity in network management and control, quickly-evolving networks require some level of autonomy enabled by device organization~\cite{Omoniwa_QL_fogRelay}.

To this end, machine learning is a promising and powerful tool to provide autonomous and effective solutions to support intelligent behaviours in UAV-enabled networks~\cite{Cui2019MARLtransactionUAV},~\cite{Wang2021}. In particular, RL has shown the capacity to address problems in UAV-enabled networks~\cite{Liu2020UAVdistributed}, \cite{Cui2019MARLtransactionUAV} -- \cite{Galkin2020UAVDeepLearning}. The goal of RL is to learn good policies for sequential decision problems, by optimizing a cumulative future reward signal~\cite{Sutton1998}. A deep reinforcement learning (DRL) approach was proposed in \cite{Liu2020UAVdistributed} and \cite{Liu2018UAV} to control a fleet of UAV-BSs, providing coverage while maintaining connectivity with each other and minimizing their energy consumption.

The decentralized approach~\cite{Liu2020UAVdistributed} was an improvement to the centralized learning approach in \cite{Liu2018UAV}, where all agents are controlled by a single actor-critic network. However, both work~\cite{Liu2020UAVdistributed} and \cite{Liu2018UAV} consider optimizing the energy efficiency of UAVs using a 2D trajectory optimization model with no consideration for interference from neighbouring UAV-BSs. Moreover, they only consider the deployment of static ground devices, thereby ignoring the impact of ground device mobility on the system performance. A clustering-based Q-learning algorithm was proposed in \cite{Liu2019UAVqlearning} to optimize the deployment of multiple UAVs providing coverage to both static and mobile devices in a pre-partitioned area. However, this approach considered the optimization of only the mean opinion score while assuming complete knowledge of the locations of all ground devices in the network. Moreover, the central cloud controller was required to periodically scan the entire network perimeter and send updates of locations of ground devices to the UAVs for local decision-making.

In a dynamic network environment, the ideal controller to UAVs assumption in \cite{Liu2019UAVqlearning} is impractical, as it requires significant information exchange between UAV and server. Moreover, it may not be possible to track user locations in a disaster scenario. To address this gap, we focus on jointly maximizing the connectivity of mobile ground devices and energy utilization of UAVs while optimizing the 3D flight trajectory of each UAV-BSs over a series of time-steps. We account for interference caused by the proximity of neighbouring UAV cells while deploying UAVs to provide wireless connectivity to both static and mobile ground devices in a given geographical space. Hence, we propose a DQLSI algorithm that relies on local observations from the UAV and the broadcast connectivity observation from neighbouring UAVs. Here, each UAV is equipped with an autonomous agent that maximizes the connectivity to both static and mobile devices while improving its energy utilization.

\vspace{-3mm}
\section{System model}
\begin{figure}[!t]
\centering
\includegraphics[width=2.8in]{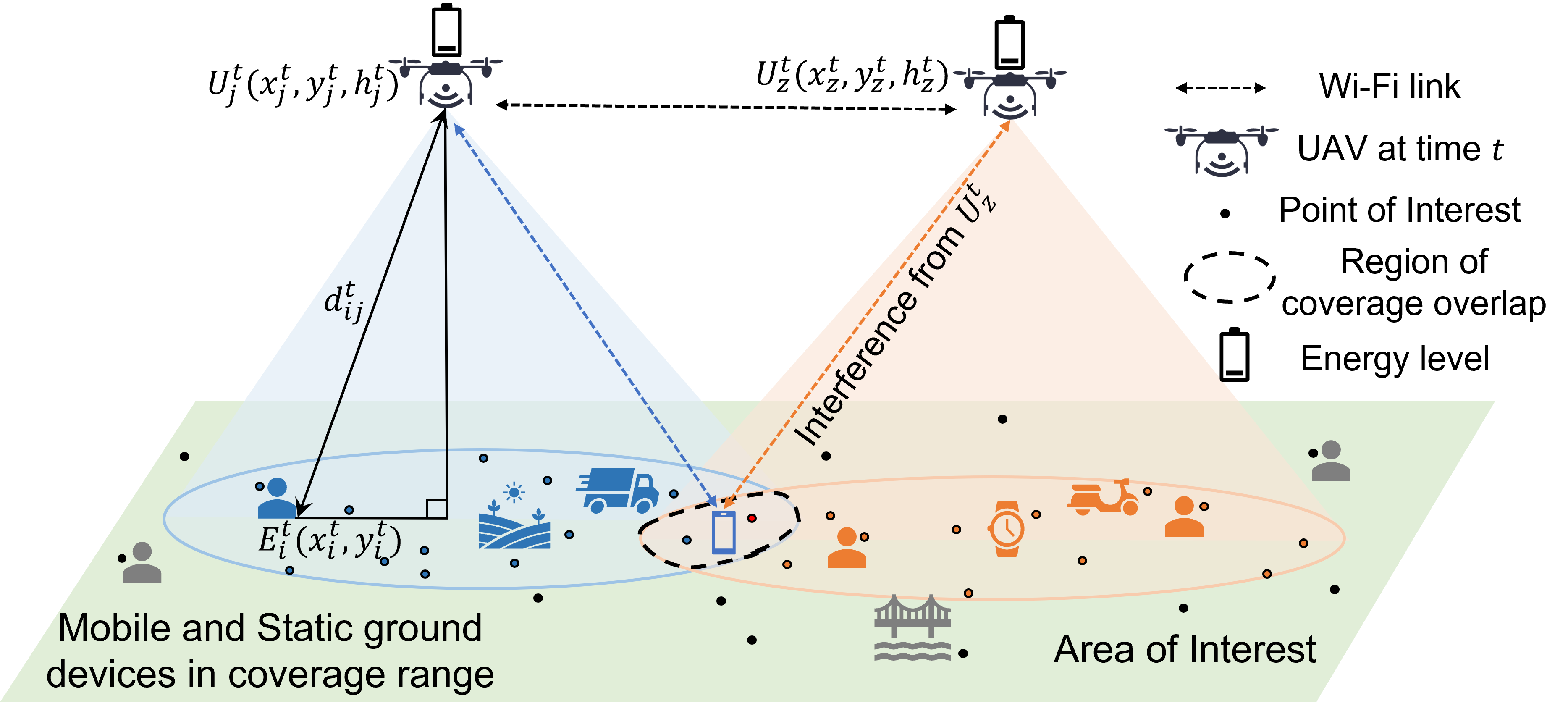}
\caption{UAV-BSs providing coverage to ground devices.}
\label{scenario}
\end{figure}

We denote the set of single-antenna wireless end-devices located within a given geographical area as $\xi$. These end-devices can either be static or mobile. The location of an end-device $i \in \xi$ at time $t$ is given by $(x_i^t, y_i^t)$. A set $N$ of quadrotor UAVs are deployed as UAV-BSs within a rectangular area to provide downlink wireless service to ground end-devices, as shown in Figure \ref{scenario}. We consider a specific scenario where existing infrastructure is damaged or unavailable due to disaster, increased network demand or failure in certain parts of the network.

\subsection{Wireless channel model}
The location of a UAV serving as UAV-BS $j \in N$ at time $t$ is given by the 3D coordinates $(x_j^t, y_j^t, h_j^t)$. We assume guaranteed LOS conditions due to the aerial positions of the UAV-BS. Each end-device in $i \in \xi$ can be served by a single UAV-BS in $j \in N$ which provides the strongest downlink signal-to-interference-plus-noise-ratio (SINR) at time $t$ for the end-device such that $\gamma_{i}^t = \argmax_{j \in N} \gamma_{ij}^t$~and $\gamma_{i}^t \geq \gamma_{th}$, where $\gamma_{ij}^t$ is the downlink SINR between the end-device at $i$ and the UAV-BS $j$ at time $t$, and $\gamma_{th}$ is the threshold SINR required for successful wireless connectivity with the end-device. The expression for the SINR at time $t$ is given below~\cite{Mozaffari2017UAV}.
\begin{equation}\label{eq:sinr}
  \gamma_{ij}^t = \frac{\eta P (d_{ij}^t)^{-\alpha}}{\Sigma_{z \in \chi_{int}} \eta P (d_{iz}^t)^{-\alpha} + \sigma^2},
\end{equation}
where $\eta$ and $\alpha$ are the attenuation factor and path loss exponent that characterizes the wireless channel, respectively. $\sigma^2$ is the power of the additive white Gaussian noise at the receiver, $d_{ij}^t$ is the distance between the $i$ and $j$ at time $t$, and expressed as $d_{ij}^t = \sqrt{(x_i^t - x_j^t)^2 + (y_i^t - y_j^t)^2 + (h_j^t)^2}$.~$\chi_{int}$ is the set of interfering UAV-BSs within the coverage range. $z$ is the index of an interfering UAV-BS in the set~$\chi_{int}$. $P$ is the transmit power of the UAV-BSs. We model the mobility of end-devices using the random walk mobility (RW) and random waypoint mobility (RWP) models~\cite{Camp_RW_2002}, i.e., the end-devices may dynamically change their positions, and the UAV-BSs can adjust their locations to provide connectivity to end-devices. Our objective here is to conserve the energy of the UAV-BSs while maximizing the provided connectivity to both static and mobile end-devices in the network.

\subsection{Connectivity model}
Network connectivity implies the existence of at least one communication path between any pair of devices~\cite{Ammari2006}. On the other hand, coverage is the area around a UAV-BS where the SINR $\gamma^t$ exceeds a pre-defined threshold $\gamma_{th}$. In this interference-limited system, coverage is affected by the SINR. We compute the connection score of a UAV-BS in location $j$ at time $t$ as,
\begin{equation}\label{eq:avgcoveragescore}
C_t^j = \sum_{\forall i \in \xi}w_t^j(i),
\end{equation}
where $w_t^j(i) \in [0, 1]$ denotes whether device $i$ is connected to UAV $j$ at time $t$. $w_t^j(i) = 1$ if $\gamma_{i}^t = \gamma_{ij}^t > \gamma_{th}$, otherwise $w_t^j(i) = 0$. In a variety of network service scenarios, including disasters, it is desirable to have nearly all ground devices connected fairly to the available UAVs. As such, we define geographical fairness using the Jain’s fairness index as~\cite{Liu2020UAVdistributed},
\begin{equation}\label{eq:JainFairness}
f_t^j = \frac{\big(\sum_{\forall j \in N}C_t(j)\big)^2}{N \sum_{\forall j \in N}C_t(j)^2}
\end{equation}

\subsection{Energy consumption model}
UAV-BSs consume energy when they are in operation, with an energy cost $e_T$. By this, the total energy consumption of a UAV-BS to complete a task is the sum of propulsion energy consumption and
communication energy consumption.
\begin{equation}\label{eq:totalenergy}
e_T = e_P + e_C,
\end{equation}
where $e_P$ is the propulsion energy to support the UAV-BSs' transition from one point to another and hovering in the air, and $e_C$ is the communication energy for wireless signal processing and data transmission. It is stated in \cite{Eom2020UAVpropulsionenergy} that $e_C$ is practically much smaller than $e_P$, i.e., $e_C \ll e_P$. Hence in this work, we ignore energy consumption from the circuits for signal processing such as channel decoders and analog-to-digital converters. A closed-form analytical propulsion power consumption model for a rotary-wing UAV-BS flying at a time $t$ is given as \cite{Zeng2019UAV},
\begin{equation}\label{eq:powertime}
P(t) = \kappa_0 \Big(1 + \frac{3V^2}{U^2_{tip}}\Big)  + \kappa_i \Big( \sqrt{1 + \frac{V^4}{4v_0^4}} + \frac{V^2}{2v_0^2}\Big)^{\frac{1}{2}} + \frac{\rho}{2} \nu s A V^3,\\
\end{equation}
where $\kappa_0$ and $\kappa_i$ are physical constants of the UAV-BS flight environment (e.g., rotor radius or weight), $U_{tip}$ is the tip speed of the rotor blade, $v_0$ is the mean rotor induced velocity during hovering, $\nu$ is the fuselage drag ratio, $s$ rotor solidity, $A$ is the rotor disc area, $V$ is the UAV-BSs' speed and $\rho$ is the air density. In particular, we take into account the basic operations of the UAV-BS, such as, hovering and acceleration. Thus, we can
obtain the average propulsion power over $T$ time-steps as $\frac{1}{T}\sum_{t=1}^{T}P(t)$, and the total consumed propulsion energy is given as,

\begin{equation}\label{eq:propenergy}
e_{P} = \delta_t \sum_{t=1}^{T}P(t),
\end{equation}
where $\delta_t$ is the duration of each time step.
\section{Decentralized Q-learning with Local Sensory Information (DQLSI)}
Deploying multiple UAV-BSs in a dynamic and quickly-evolving network makes energy estimation and planning complex and difficult. As such, we propose a decentralized multi-agent reinforcement learning-based approach \cite{Busoniu2008} to improve the energy utilization of multiple UAV-BSs, while maximizing the connectivity of both static and mobile ground end-devices. We consider a framework of cooperative multi-agent system (MAS), where agents learn locally and independently. In such environments, the agents share the common interest of maximizing wireless connectivity while improving the energy utilization within the network. In this work, we focus on agents with local observability called independent learners (IL), since the assumption of joint action observability is unrealistic without central/global knowledge~\cite{Busoniu2008}.
The Q-learning update for agent $i$ is given as,
\begin{equation}\label{eq:Decentralqlearningeqn}
Q_i(s_i, a_i) \gets (1-\alpha) Q_i(s_i, a_i) + \alpha \Big[ r_i + \gamma \max_{a_i'}  Q_i(s_i', a_i') \Big],
\end{equation}
where $s_i$ is the present local state observed by agent $i$, $s_i'$ is the new local state observed by agent $i$, $a_i$ is the action taken by agent $i$, $r_i$ is the reward received by agent $i$ in that time step, $\alpha$ is the learning rate and $\gamma \in [0, 1]$ is the discount factor.
In our multi-UAV-BS system, we consider a DQLSI algorithm as shown in Algorithm \ref{Algorithm_QL} with strict local observability suitable when there is limited or no access to cellular infrastructure due to disaster. We assume that each UAV-BS is equipped with an autonomous agent which takes an action and in turn, receives a reward and makes a transition to a new state.
\begin{algorithm}
\footnotesize
\caption{Decentralized Q-learning with Local Sensory Information (DQLSI) for Agent $i$}\label{Algorithm_QL}
\begin{algorithmic}[1]
\State \textbf{Initialize:}
\ForAll{$a \in A_i$ and~$s \in S$}:
\State $Q_{i, max}(s, a)$ $\gets$ 0,~${\pi_{i}(s, a)}$ arbitrarily
\State $s$ $\gets$ initial state
\State 10000 $\gets$ maxStep
\State \texttt{\%\% An episode ends when \emph{goal} is Reached or UAV-BS \emph{dies} or maxStep is reached}
\While{\emph{goal} not Reached and Agent \emph{alive} and maxStep not reached}
\State \emph{s} $\leftarrow$ MapLocalObservationToState(\emph{Env})
\State \texttt{\%\%From s select a according to $\epsilon$-greedy method based on $\pi_i$}
\State \emph{a} $\leftarrow$ QLearning.SelectAction(\emph{s})
\State \texttt{\%\% AgentExecutesActionInState}
\State  \emph{a}.execute(\emph{Env})
\If { \emph{a}.execute(\emph{Env}) is True}
\State \texttt{\%\% MapToNewState}
\State Env.UAV3Dposition \cite{Liu2019UAVqlearning}
\State ~~~\textit{Observation}.ConnectionScore (\ref{eq:avgcoveragescore})
\State ~~~\textit{Observation}.EnergyConsumed (\ref{eq:propenergy})
\State ~~~\textit{Observation}.GeographicalAreaCovered
\State Env.UAVneighbourProximity
\State ~~~\textit{Observation}.ConnectionSensation
\EndIf
\State \texttt{\%\%Observe reward $r$ and next state $s'$}
\State UpdateQLearningProcedure() (\ref{eq:Decentralqlearningeqn})
\EndWhile
\State \textbf{endwhile}
\EndFor
\end{algorithmic}
\end{algorithm}
We explicitly define the states, actions, and reward function of our agent.
\subsection{States}
We consider a combination of the three-dimensional (3D) position of the UAV-BS~\cite{Liu2019UAVqlearning} and the distance from neighbouring UAV-BSs. This helps to inform the agent of its position and that of neighbouring agents in each time-step. The agent observes its connectivity to ground devices, energy consumption, and the geographical ground area covered at each position in space. In addition, each agent can locally observe the broadcast of its neighbour's connection score. However, the communication cost incurred by each sensory-exchanging agent per step is bounded by $(N - 1) \times E$ (Refer to Case-study 2, \cite{Ming_MARL_1993}), where $N$ is the number of UAV-BSs within that locality,  $E$ is the number of bits needed to represent the connection sensation observed by an agent. The agent then discretizes the continuous state space emanating from the environment.
\vspace{-2mm}
\subsection{Actions}
\vspace{-1mm}
At each time-step~$t \in T$, each UAV-BS takes an action~$a_t \in A$. We discretize the agent's actions~\cite{Liu2019UAVqlearning}:~Up, Down, Forward, Backward, Left, Right and Stationary. Our rationale for discretizing the action space was to ensure quick adaptability and convergence of the agents considering the limited battery capacity of UAVs and the need to quickly deploy high quality service to ground devices in emergency scenarios.
\vspace{-1mm}
\subsection{Reward}
The goal of each learning agent is to learn a policy that maximizes the connectivity of ground devices while improving its energy utilization. The reward assigned to an agent $i$ in each time-step $t \in T$ is given in as,
\begin{equation}\label{eqnreward}
\mathcal{R}^i =
\begin{cases}
  ~~\mho + \frac{e_{t - 1}^i - e_{t}^i}{e_{t}^i + ~e_{t - 1}^i} + 1, & \text{if}\ C_{t}^i > ~C_{t - 1}^i\\
  ~~\mho + \frac{e_{t - 1}^i - e_{t}^i}{e_{t}^i + ~e_{t - 1}^i}, & \text{if}\ C_{t}^i = ~C_{t - 1}^i\\
  ~~\mho + \frac{e_{t - 1}^i - e_{t}^i}{e_{t}^i + ~e_{t - 1}^i} -1, & \text{otherwise,}
\end{cases}
\end{equation}
where $C_{t}^i$ and $C_{t - 1}^i$ are the connection score and $e_{t}^i$ and $e_{t - 1}^i$ are the average energy consumed by agent $i$ in present and previous time-step, respectively. $\mho$ is a cooperative factor which evaluates the connectivity among neighbouring UAV-BSs.~$\mho = +1$ if $C_{t}^o > ~C_{t - 1}^o$, otherwise $\mho = -1$, where $C_{k}^o = C_{k}^i + \sum^N_{-i} C_{k}^{-i}$, is the overall connectivity score in the locality, and  $\sum^N_{-i} C_{k}^{-i}$ is the connection score from neighbouring UAV-BSs in $k^{th}$ time-step.
\vspace{-3mm}
\section{Evaluation, Results and Analysis}
\begin{figure}[!t]
\centering
\includegraphics[width=3.0in]{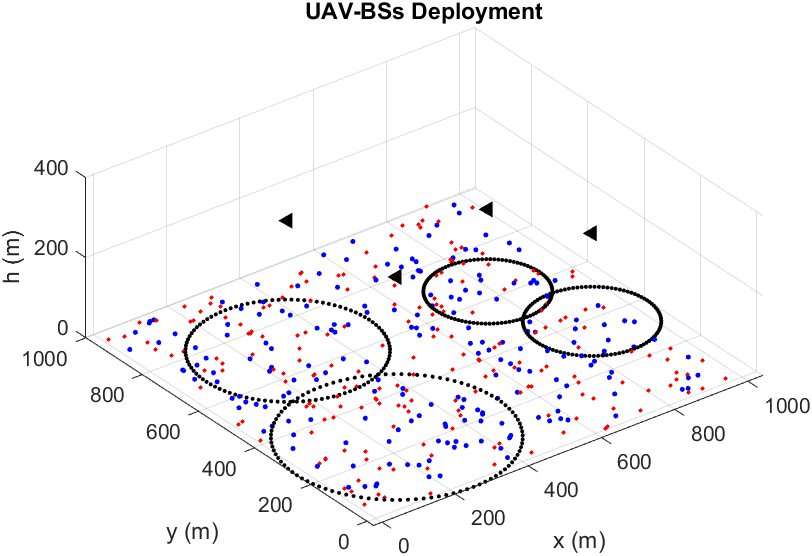}
\caption{Simulation snapshot of four UAV-BSs providing wireless coverage to 200 static (blue) and 200 mobile (red) evenly-distributed ground devices.}
\label{snapshot}
\end{figure}

\vspace{-1mm}
In our experiments, we consider settings with static and dynamic ground devices. In both, four UAV-BSs were deployed in a 1~$km^2$ area, each with its initial starting point at the start of an episode. The mobility step size for each UAV-BS is 20 meters. In the static setting, we deploy 400 randomly distributed static ground devices that are not uniformly distributed in the area. We deploy 400 devices (200 static and 200 mobile devices) in the dynamic setting as seen in Figure \ref{snapshot}. The random walk mobility (RW) and random waypoint mobility (RWP) models were used to simulate the unpredictable movement of ground devices in a campus setting~\cite{Camp_RW_2002}. We assume a maximum connection limit of 150 active ground devices per UAV-BS~\cite{Mozaffari2017UAV}.

We consider the cluster-based Q-learning (CQL)~\cite{Liu2019UAVqlearning}, iterative search (IS)~\cite{Mozaffari2017UAV}, and exhaustive (brute-force) search (ES) as baselines to our proposed DQLSI algorithm. Using the same energy parameters given in (\ref{eq:propenergy})~\cite{Zeng2019UAV} and \cite{Hwang_2018}, we consider the total energy consumed by the UAV-BS over the range of time-steps $T$ as a metric expressed in Joules. We account for interference from neighbouring UAVs (i.e., taking into account the possibility of coverage overlap by multiple UAV-BSs), hence, the number of ground devices connected to a UAV-BS is expressed as a percentage of the total number of ground devices. Lastly, we observed the geographical fairness using (\ref{eq:JainFairness}) and area covered by UAV-BS, which estimates in square kilometers the actual ground coverage by all UAV-BSs. We used statistical results from 20 independent runs, gathering 5 episodes where the DQLSI agent was sufficiently trained in each run of experiment.
\subsection{Static setting}
\begin{figure}[ht]
\begin{minipage}[b]{0.45\linewidth}
\centering
\subfloat[Subfigure 1 list of figures text][Accumulated rewards per agent for 100 learning episodes.]{
\includegraphics[width=\textwidth]{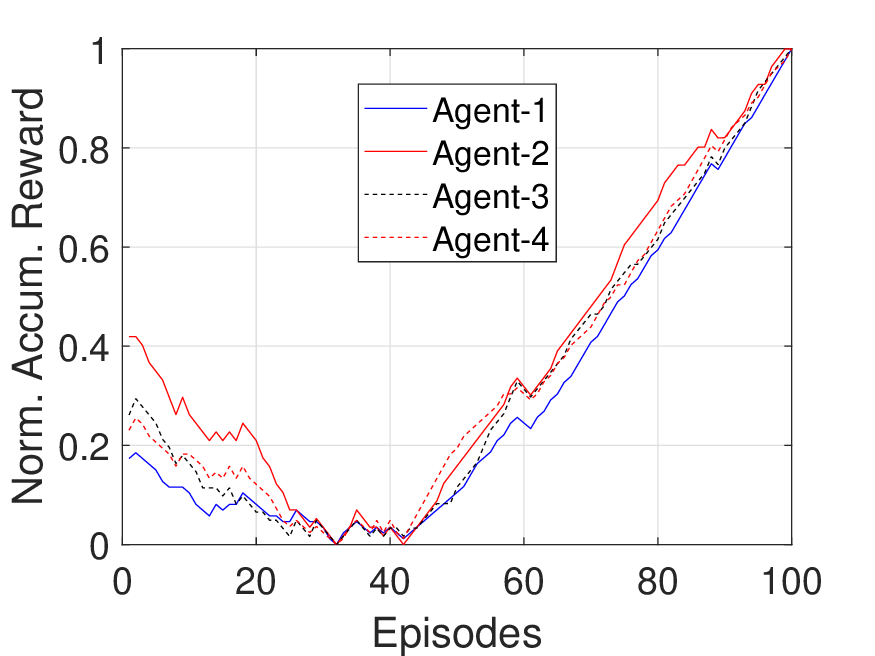}
\label{fig:subfig1_static_reward}}
\end{minipage}
\hspace{0.5cm}
\begin{minipage}[b]{0.45\linewidth}
\centering
\subfloat[Subfigure 2 list of figures text][Energy consumed per agent.]{\includegraphics[width=\textwidth]{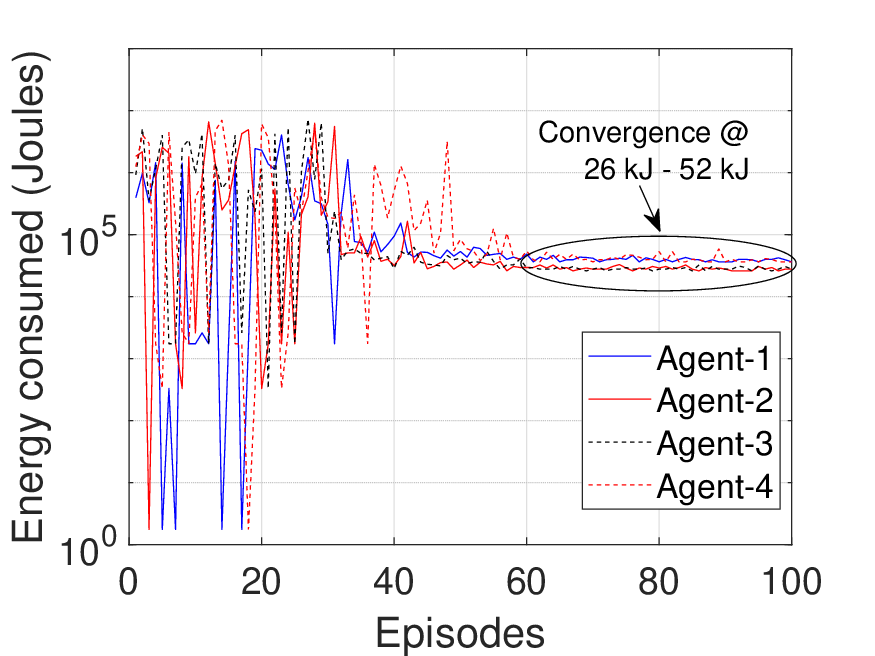}
\label{fig:subfig2_static_energy}}
\end{minipage}
\begin{minipage}[b]{0.45\linewidth}
\centering
\subfloat[Subfigure 3 list of figures text][Number of connected ground device per agent in the network.]{
\includegraphics[width=\textwidth]{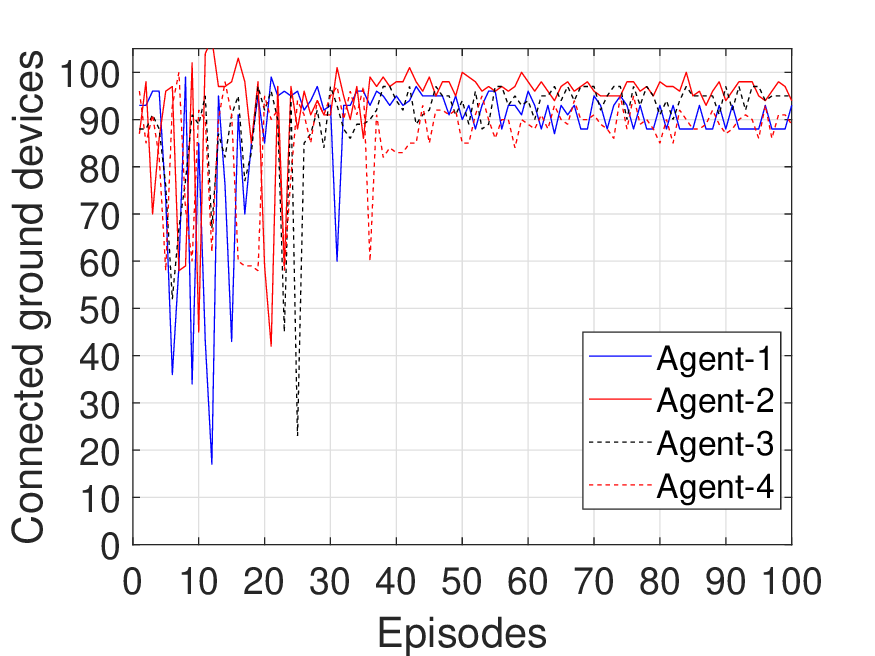}
\label{fig:subfig3_static_coverage}}
\end{minipage}
\hspace{0.5cm}
\begin{minipage}[b]{0.45\linewidth}
\centering
\subfloat[Subfigure 6 list of figures text][Geographical fairness and area covered by multi-UAV-BS.]{\includegraphics[width=\textwidth]{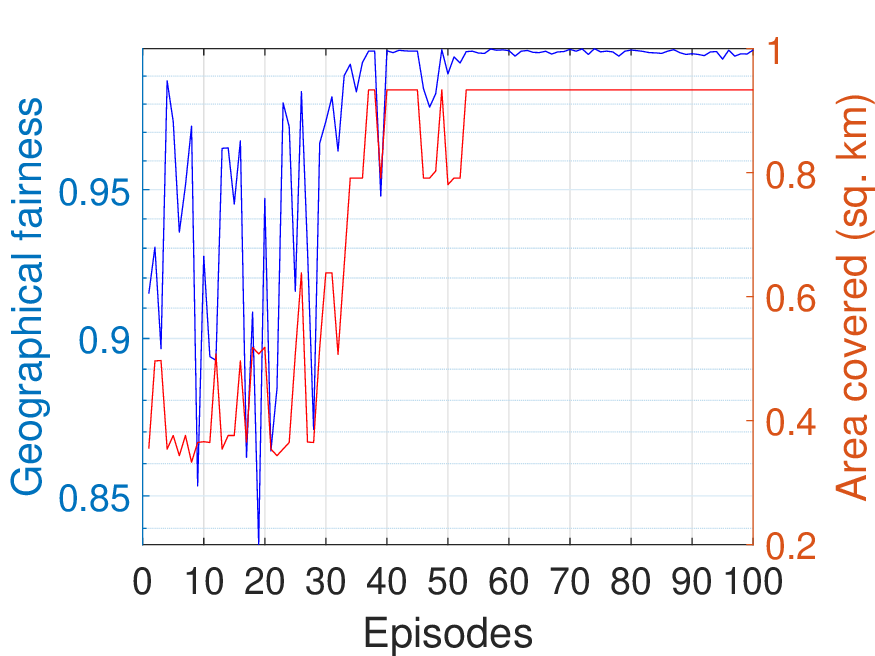}
\label{fig:subfig6_static_geog}}
\end{minipage}
\caption{Four UAV-BSs providing connectivity in static environments with 400 randomly distributed static ground devices.}
\label{fig: fourUAV_static}
\end{figure}

Figure \ref{fig: fourUAV_static} shows performance metrics of each agent when the proposed DQLSI algorithm is used in a static setting. As seen in Figure \ref{fig:subfig1_static_reward}, all four agents maximize their accumulated reward. After the $60^{th}$ episode, we observe significant convergence in the energy consumed by the UAVs, within the range of about 26 kJ - 52 kJ as seen in Figure \ref{fig:subfig2_static_energy}. Figure \ref{fig:subfig3_static_coverage} shows a balance in the connection load across the four UAVs, ranging between 85~-~100 connected ground devices per UAV. The total number of connected ground devices for all UAVs ranges between 365~-~382. Figure \ref{fig:subfig6_static_geog} shows convergence in the geographical area covered by the UAVs after the $40^{th}$ episode.

\subsection{Dynamic setting with even randomly-distributed ground devices}
\begin{figure}[ht]
\begin{minipage}[b]{0.45\linewidth}
\centering
\subfloat[Subfigure 1 list of figures text][Accumulated rewards per agent for 100 learning episodes.]{
\includegraphics[width=\textwidth]{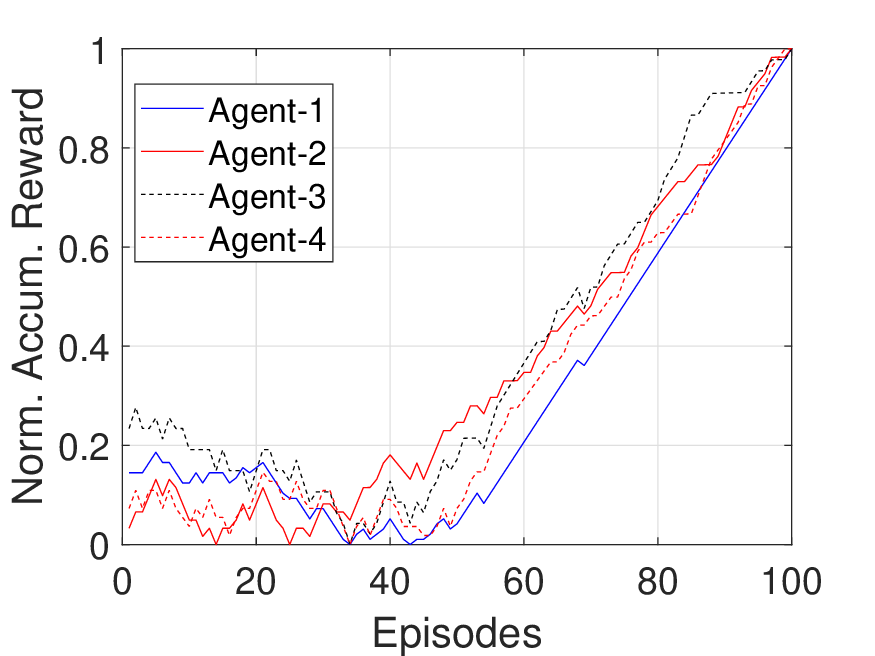}
\label{fig:subfig1_dyn_reward}}
\end{minipage}
\hspace{0.5cm}
\begin{minipage}[b]{0.45\linewidth}
\centering
\subfloat[Subfigure 2 list of figures text][Energy consumed per agent.]{\includegraphics[width=\textwidth]{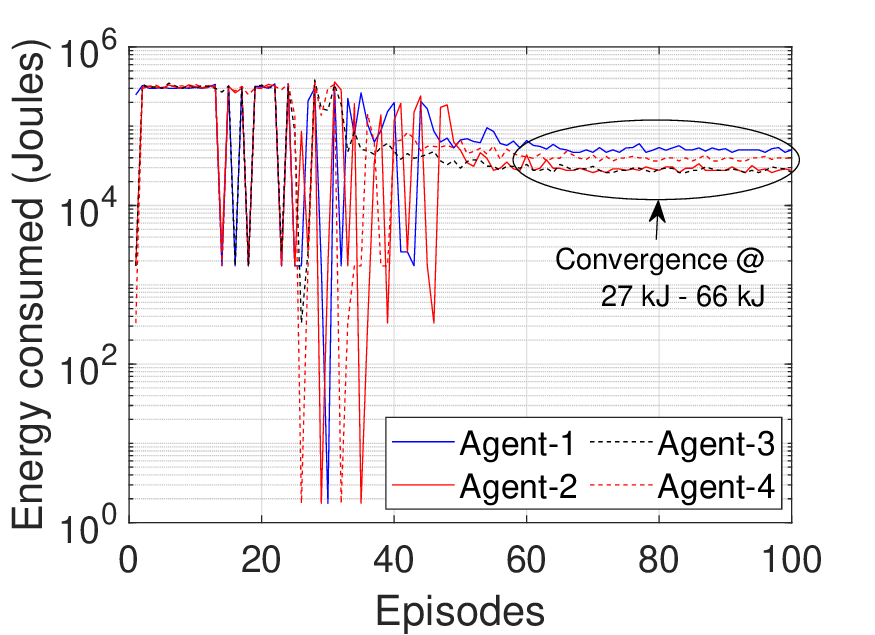}
\label{fig:subfig2_dyn_energy}}
\end{minipage}
\begin{minipage}[b]{0.45\linewidth}
\centering
\subfloat[Subfigure 3 list of figures text][Number of connected ground device per agent in the network.]{
\includegraphics[width=\textwidth]{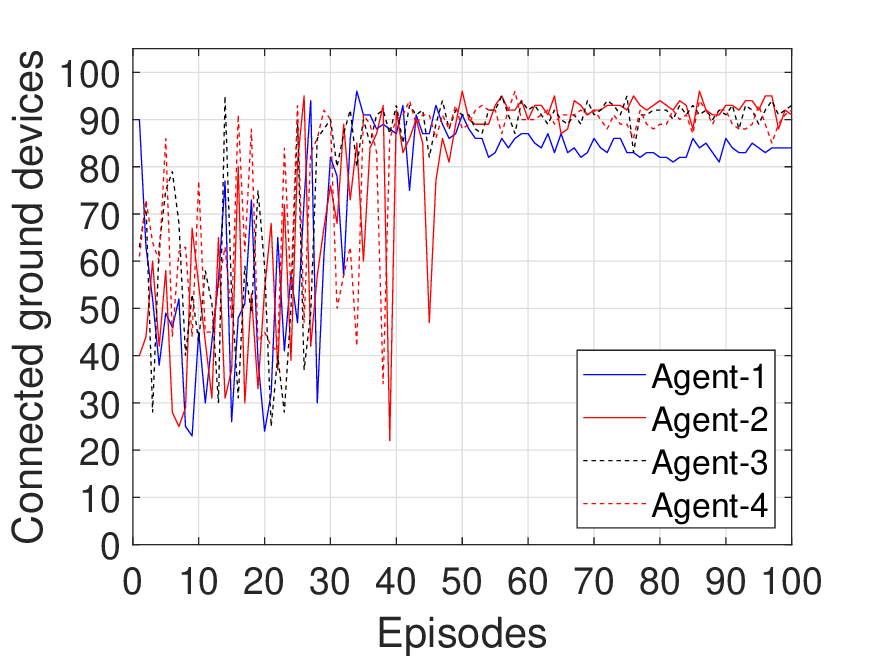}
\label{fig:subfig3_dyn_coverage}}
\end{minipage}
\hspace{0.5cm}
\begin{minipage}[b]{0.45\linewidth}
\centering
\subfloat[Subfigure 4 list of figures text][Geographical fairness and area covered by multi-UAV-BS.]{\includegraphics[width=\textwidth]{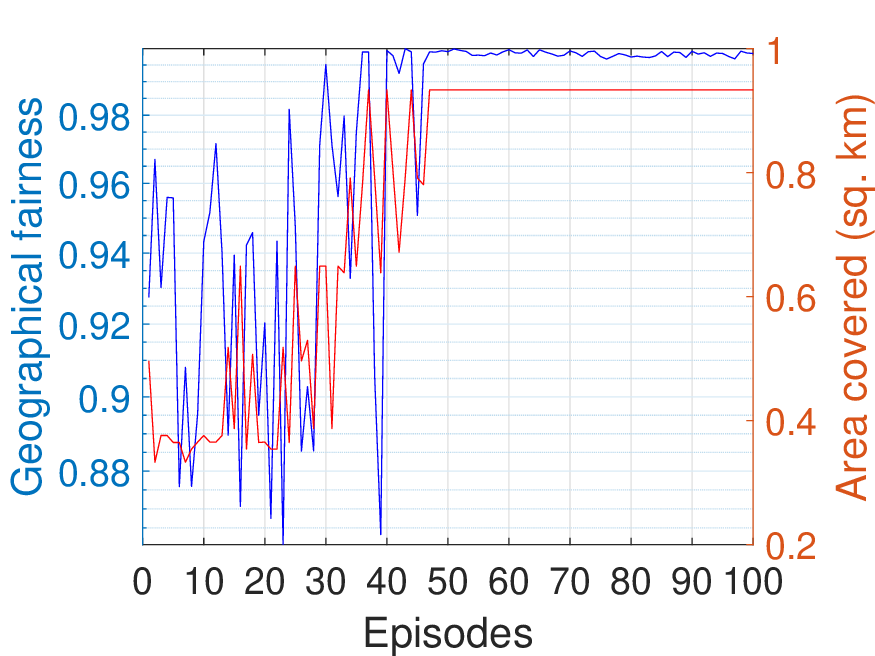}
\label{fig:subfig6_dyn_geog_overall}}
\end{minipage}
\caption{Four UAV-BSs providing connectivity in dynamic environments with 200 randomly distributed static ground devices and 200 mobile RW-based (random walk mobility model) devices.}
\label{fig: fourUAV_dyn}
\end{figure}

\begin{figure}[ht]
\begin{minipage}[b]{0.45\linewidth}
\centering
\subfloat[Subfigure 1 list of figures text][Accumulated rewards per agent for 100 learning episodes.]{
\includegraphics[width=\textwidth]{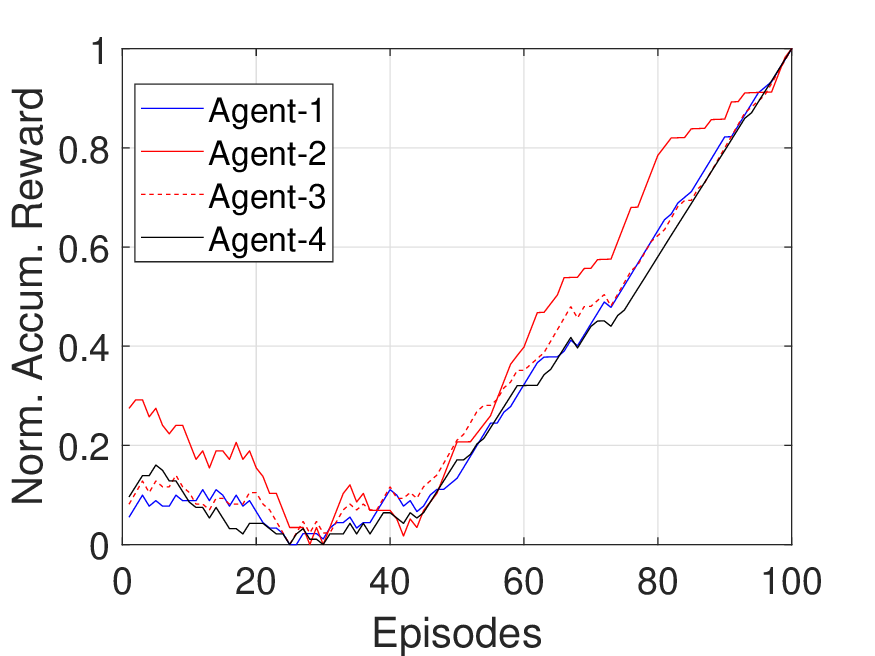}
\label{fig:subfig1_dyn_reward_rwp}}
\end{minipage}
\hspace{0.5cm}
\begin{minipage}[b]{0.45\linewidth}
\centering
\subfloat[Subfigure 2 list of figures text][Energy consumed per agent.]{\includegraphics[width=\textwidth]{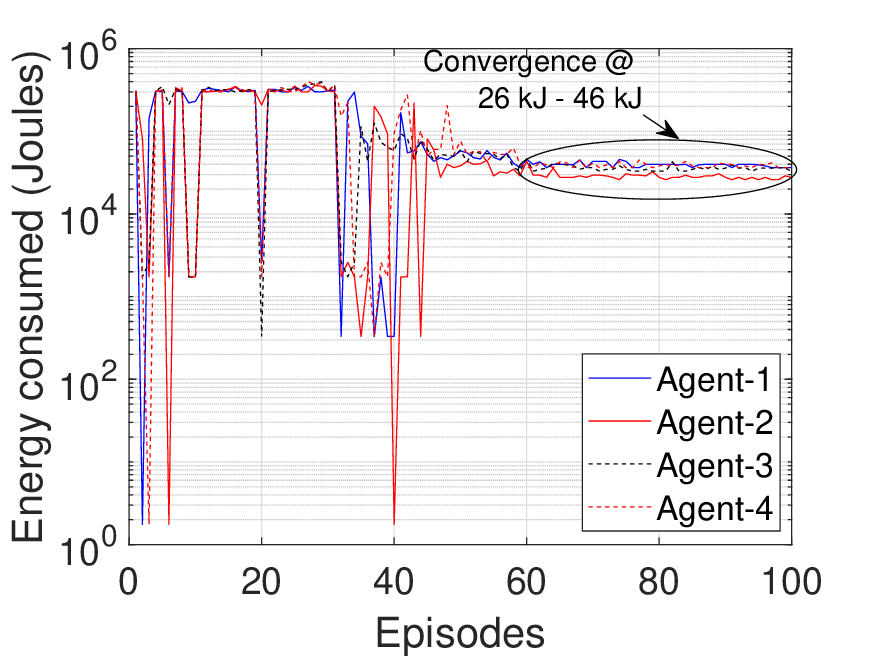}
\label{fig:subfig2_dyn_energy_rwp}}
\end{minipage}
\begin{minipage}[b]{0.45\linewidth}
\centering
\subfloat[Subfigure 3 list of figures text][Number of connected ground device per agent in the network.]{
\includegraphics[width=\textwidth]{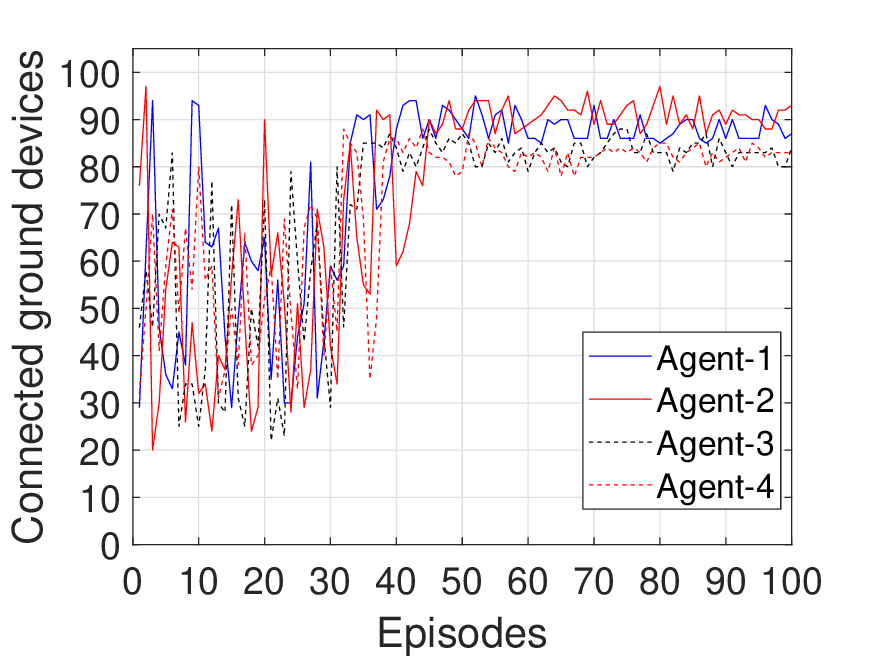}
\label{fig:subfig3_dyn_coverage_rwp}}
\end{minipage}
\hspace{0.5cm}
\begin{minipage}[b]{0.45\linewidth}
\centering
\subfloat[Subfigure 4 list of figures text][Geographical fairness and area covered by multi-UAV-BS.]{\includegraphics[width=\textwidth]{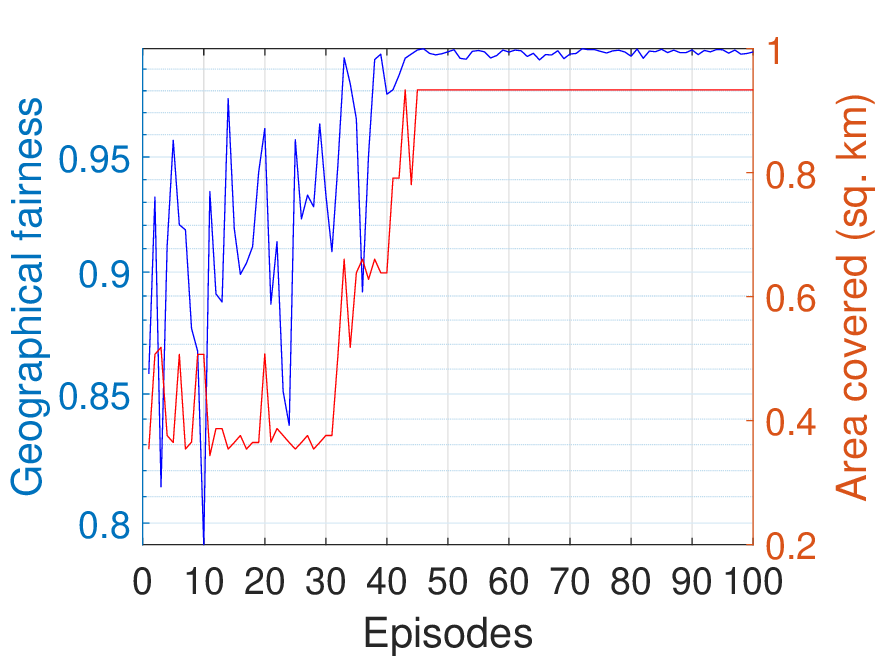}
\label{fig:subfig6_dyn_geog_overall_rwp}}
\end{minipage}
\caption{Four UAV-BSs providing connectivity in dynamic environments with 200 randomly distributed static ground devices and 200 mobile RWP-based (random waypoint mobility model) devices.}
\label{fig: fourUAV_dyn_rwp}
\end{figure}

Figure \ref{fig: fourUAV_dyn} and Figure \ref{fig: fourUAV_dyn_rwp} shows the performance metrics of each agent when the proposed DQLSI algorithm is applied in an even randomly-distributed (DQLSI-D) dynamic setting under the RW and RWP models, respectively. As seen in Figure \ref{fig:subfig1_dyn_reward} and Figure \ref{fig:subfig1_dyn_reward_rwp}, all four agents maximize their accumulated reward. Both Figure \ref{fig:subfig2_dyn_energy} and Figure \ref{fig:subfig2_dyn_energy_rwp} shows convergence in the energy consumed by the UAVs after the $60^{th}$ episode, within the range of 27 kJ - 66 kJ and 26 kJ - 46 kJ, respectively. Despite the network dynamics, Figure \ref{fig:subfig3_dyn_coverage} and Figure \ref{fig:subfig3_dyn_coverage_rwp} show balance in the connection load after the $50^{th}$ episode, however, in Figure \ref{fig:subfig3_dyn_coverage} Agent~1 covered slightly fewer number of devices than other UAVs in the network. The total number of connected ground devices by all UAVs ranges between 347~-~365 and 339~-~354 in the dynamic settings using the RW and RWP models, respectively. Figure \ref{fig:subfig6_dyn_geog_overall} and Figure \ref{fig:subfig6_dyn_geog_overall_rwp} show good convergence in both the geographical fairness and the area covered by the UAVs after the $50^{th}$ episode.

\subsection{Dynamic setting with uneven randomly-distributed ground devices}
\begin{figure}[ht]
\begin{minipage}[b]{0.45\linewidth}
\centering
\subfloat[Subfigure 1 list of figures text][Accumulated rewards per agent for 100 learning episodes.]{
\includegraphics[width=\textwidth]{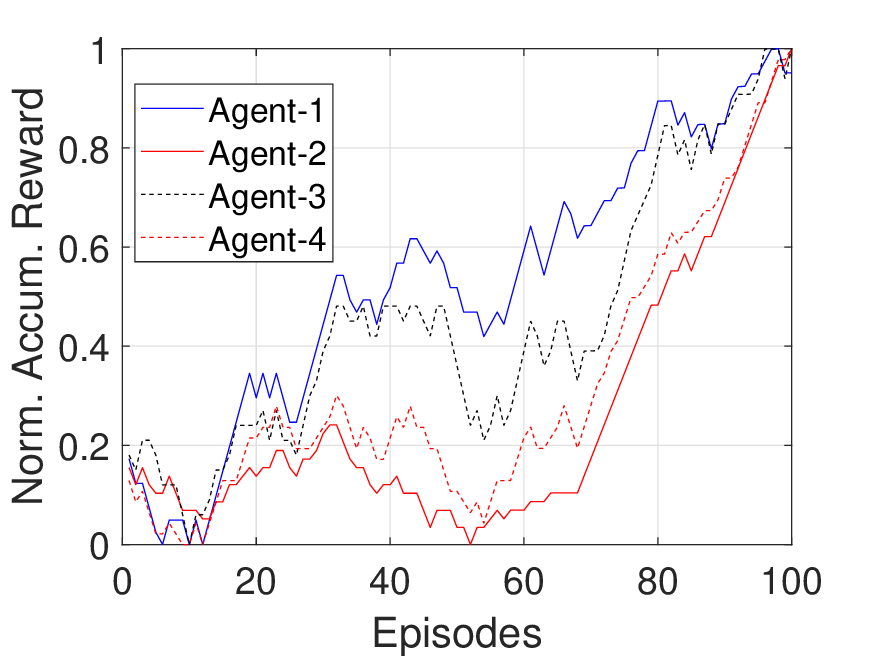}
\label{fig:subfig1_uneven_reward}}
\end{minipage}
\hspace{0.5cm}
\begin{minipage}[b]{0.45\linewidth}
\centering
\subfloat[Subfigure 2 list of figures text][Energy consumed per agent.]{\includegraphics[width=\textwidth]{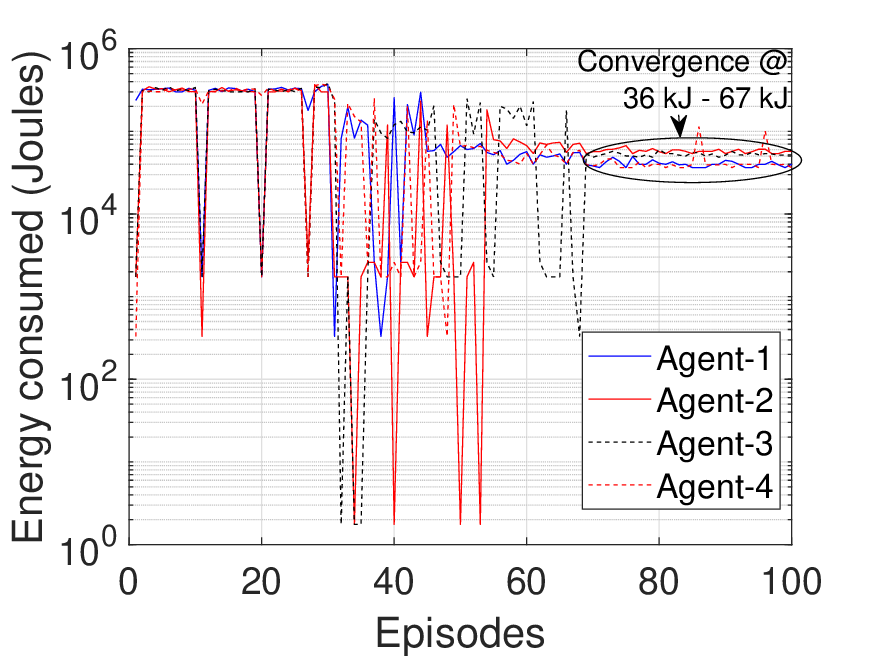}
\label{fig:subfig2_uneven_energy}}
\end{minipage}
\begin{minipage}[b]{0.45\linewidth}
\centering
\subfloat[Subfigure 3 list of figures text][Number of connected ground device per agent in the network.]{
\includegraphics[width=\textwidth]{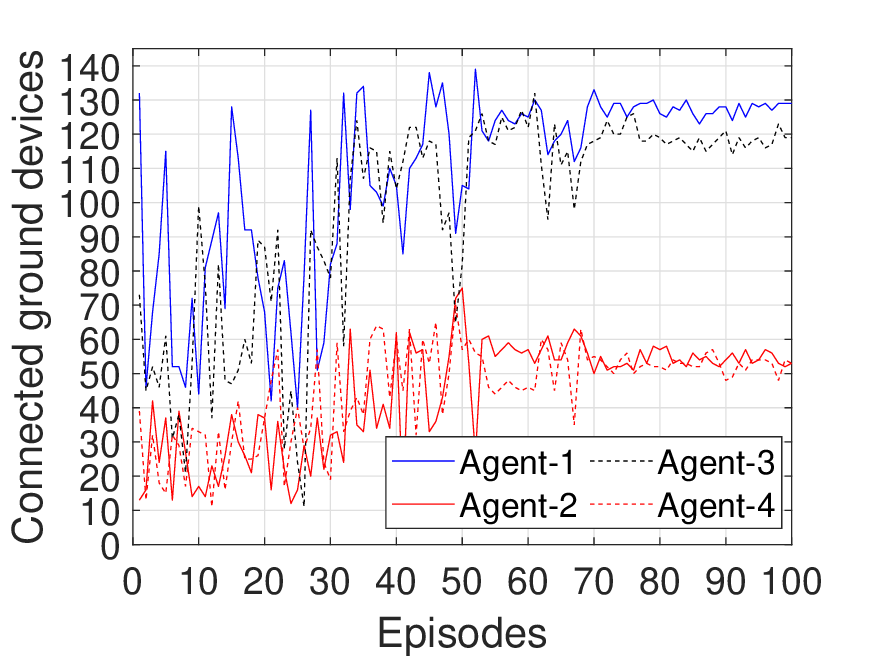}
\label{fig:subfig3_uneven_coverage}}
\end{minipage}
\hspace{0.5cm}
\begin{minipage}[b]{0.45\linewidth}
\centering
\subfloat[Subfigure 4 list of figures text][Geographical fairness and area covered by multi-UAV-BS.]{\includegraphics[width=\textwidth]{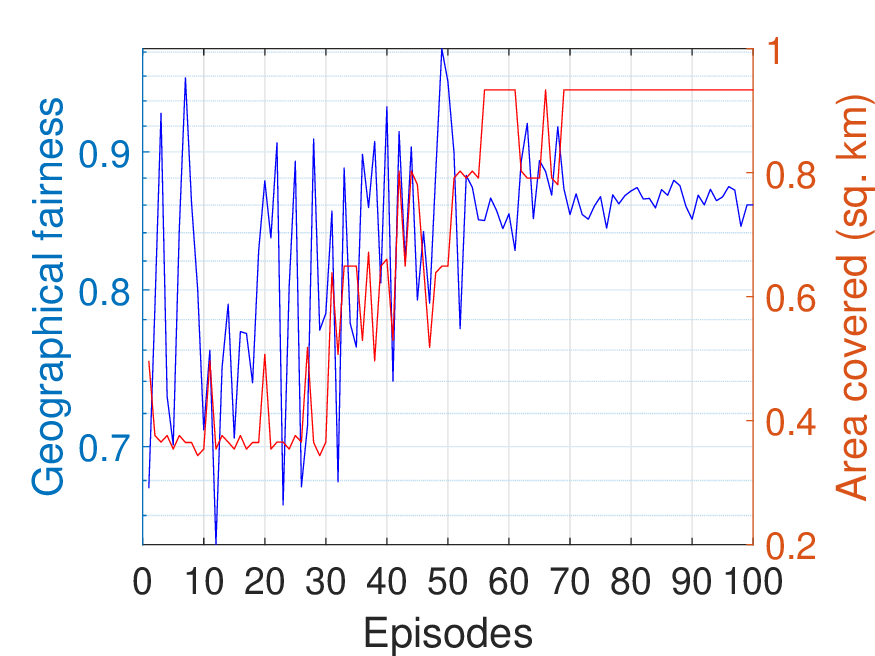}
\label{fig:subfig6_uneven_geog_overall}}
\end{minipage}
\caption{Four UAV-BSs providing connectivity in an uneven dynamic environments with 200 randomly distributed static ground devices and 200 mobile RW-based (random walk mobility model) devices.}
\label{fig: fourUAV_uneven}
\end{figure}

\begin{figure}[ht]
\begin{minipage}[b]{0.45\linewidth}
\centering
\subfloat[Subfigure 1 list of figures text][Accumulated rewards per agent for 100 learning episodes.]{
\includegraphics[width=\textwidth]{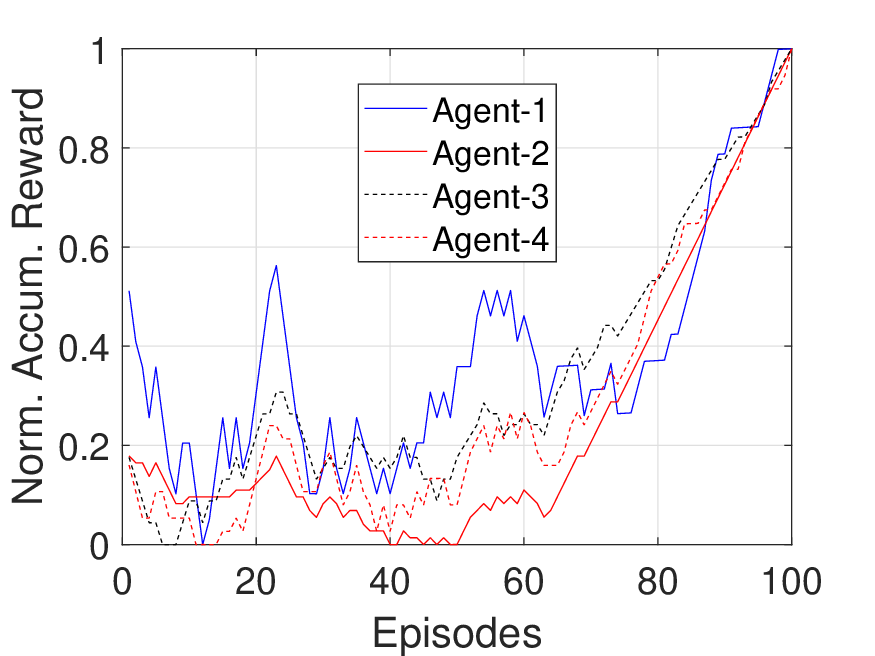}
\label{fig:subfig1_uneven_reward_rwp}}
\end{minipage}
\hspace{0.5cm}
\begin{minipage}[b]{0.45\linewidth}
\centering
\subfloat[Subfigure 2 list of figures text][Energy consumed per agent.]{\includegraphics[width=\textwidth]{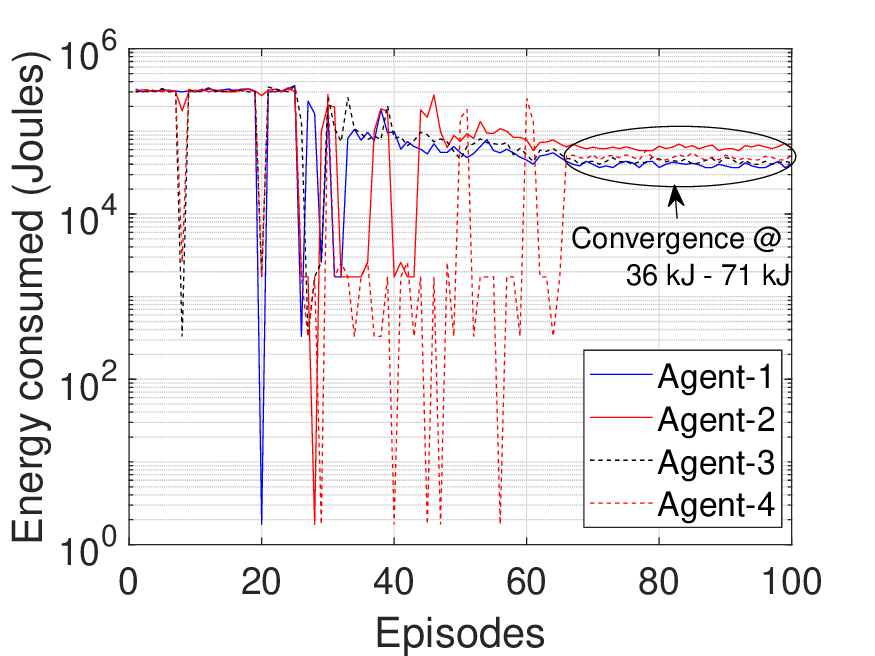}
\label{fig:subfig2_uneven_energy_rwp}}
\end{minipage}
\begin{minipage}[b]{0.45\linewidth}
\centering
\subfloat[Subfigure 3 list of figures text][Number of connected ground device per agent in the network.]{
\includegraphics[width=\textwidth]{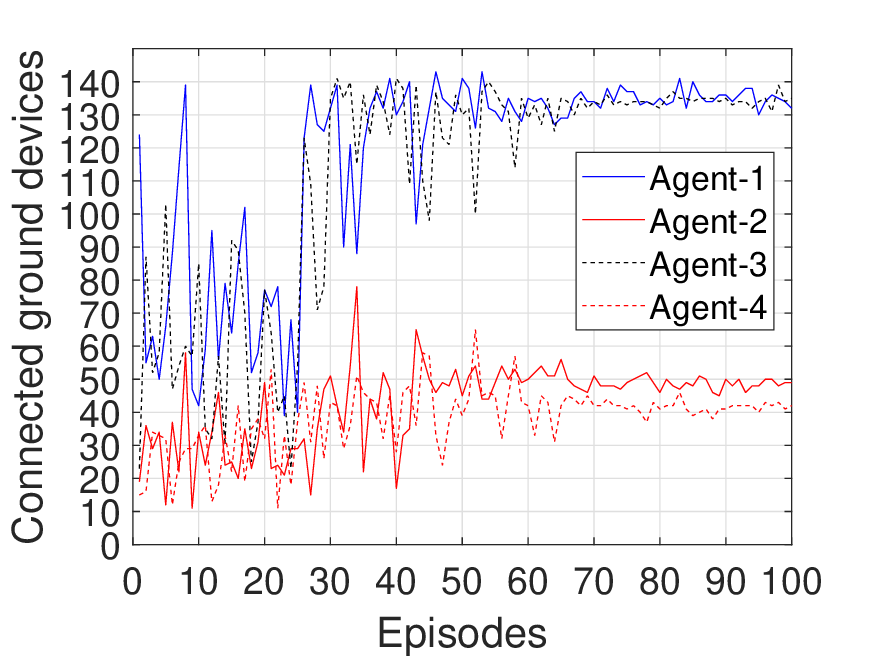}
\label{fig:subfig3_uneven_coverage_rwp}}
\end{minipage}
\hspace{0.5cm}
\begin{minipage}[b]{0.45\linewidth}
\centering
\subfloat[Subfigure 4 list of figures text][Geographical fairness and area covered by multi-UAV-BS.]{\includegraphics[width=\textwidth]{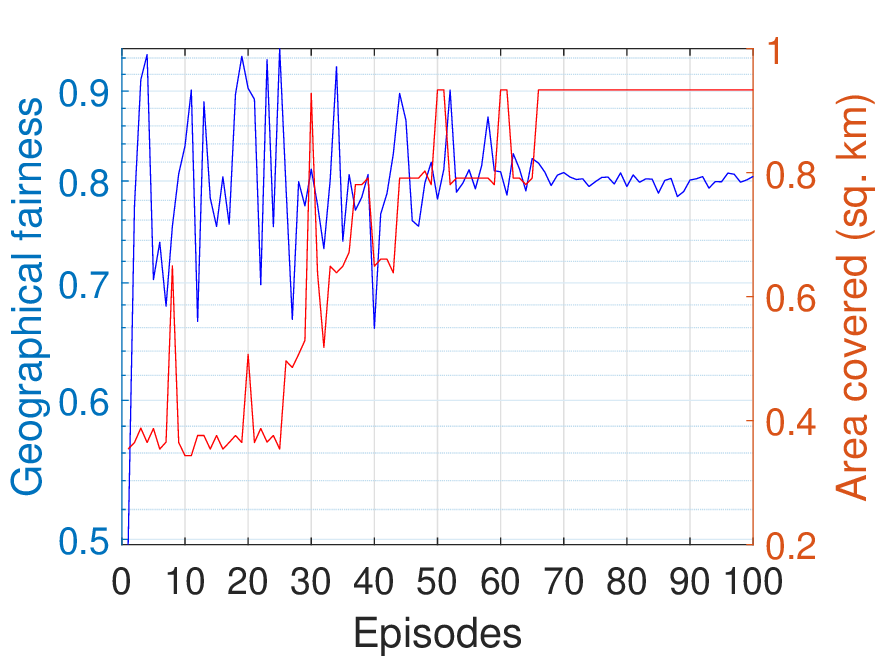}
\label{fig:subfig6_uneven_geog_overall_rwp}}
\end{minipage}
\caption{Four UAV-BSs providing connectivity in an uneven dynamic environments with 200 randomly distributed static ground devices and 200 mobile RWP-based (random waypoint mobility model) devices.}
\label{fig: fourUAV_uneven_rwp}
\end{figure}

Figure \ref{fig: fourUAV_uneven} and Figure \ref{fig: fourUAV_uneven_rwp} show the performance metrics of each agent when the proposed DQLSI algorithm is applied in an uneven randomly-distributed (DQLSI-UD) dynamic setting under the RW and RWP models, respectively. As seen in Figure \ref{fig:subfig1_uneven_reward} and Figure \ref{fig:subfig1_uneven_reward_rwp}, all four agents maximize their accumulated reward.
Both Figure \ref{fig:subfig2_uneven_energy} and Figure \ref{fig:subfig2_uneven_energy_rwp} show convergence in the energy consumed by the UAVs after the $70^{th}$ episode, within the range of 36 kJ - 67 kJ and 36 kJ - 71 kJ, respectively. In Figure \ref{fig:subfig3_uneven_coverage} and Figure \ref{fig:subfig3_uneven_coverage_rwp}, due to the uneven distribution of ground devices, Agent~1 and Agent~3 connected to twice as many devices as Agent 2 and Agent 4, possibly influenced by their initial starting position. The total number of connected ground devices by all UAVs ranges between 343~-~360 and 352~-~369 in the uneven dynamic setting using the RW and RWP models, respectively. Figure \ref{fig:subfig6_uneven_geog_overall} and Figure \ref{fig:subfig6_uneven_geog_overall_rwp} show convergence in both the geographical fairness and the area covered by the UAVs after the $70^{th}$ episode. However, the fairness index significantly dropped in uneven dynamic setting compared to the even dynamic setting.

Overall, we observe good connectivity when the proposed DQLSI algorithm is applied to all 3 scenarios. However, there was a significant drop in the number of connected ground devices in the dynamic settings as compared to the static setting, due to the quickly-evolving network topology, emphasizing the importance of developing approaches such as ours which take into account mobile devices.

\subsection{Comparative analysis}
\begin{figure}[ht]
\centering
\includegraphics[width=3.0in]{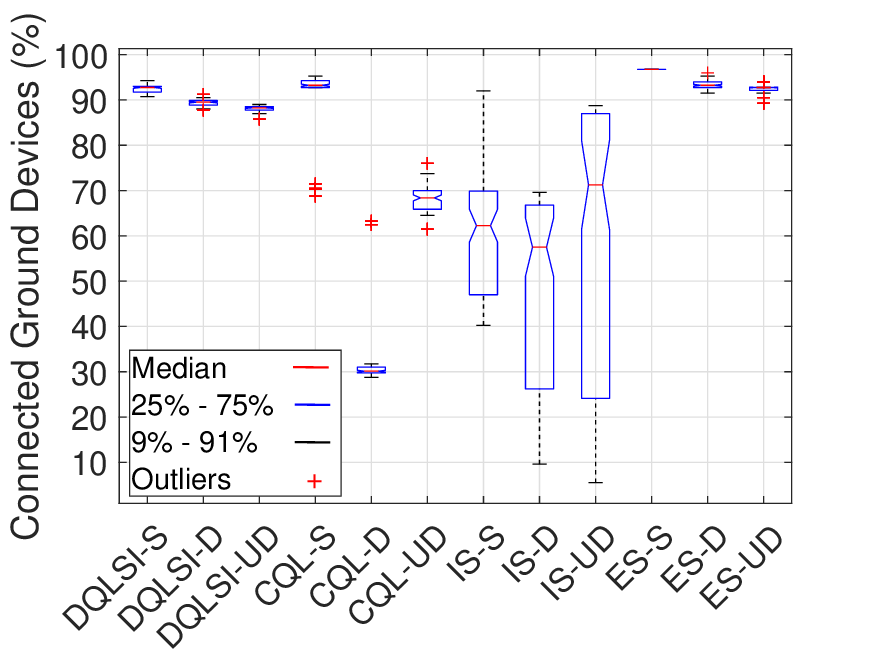}
\caption{Comparison of ground device connections in a 4-agent setting vs. baselines in both static and dynamic environments.}
\label{Fig:Cov_results}
\end{figure}

\begin{figure}[ht]
\centering
\includegraphics[width=3.0in]{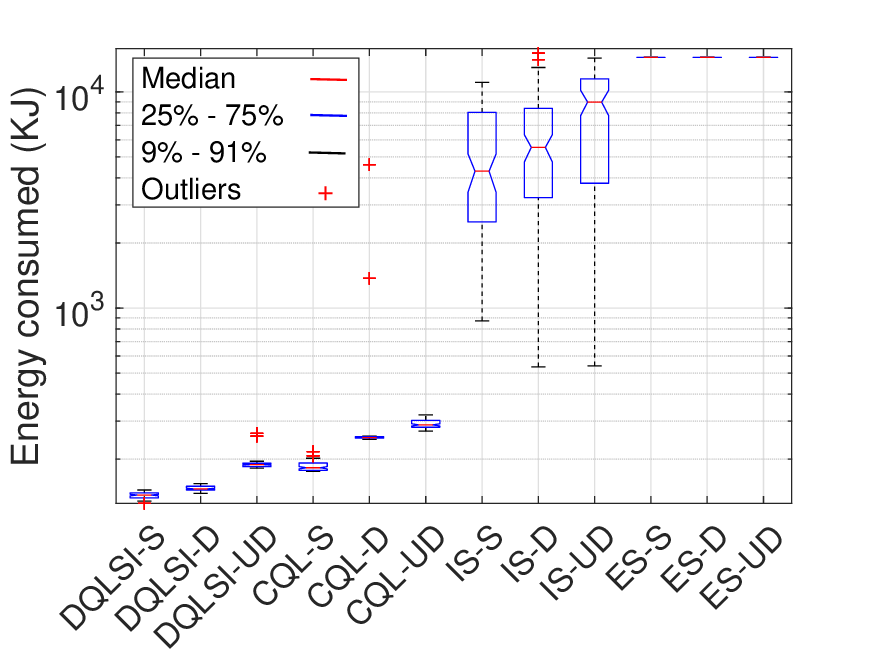}
\caption{Comparison of energy consumed in a 4-agent setting vs. baselines in both static and dynamic environments.}
\label{Fig:energy_results}
\end{figure}

We compare our proposed DQLSI approach with baselines. Figure \ref{Fig:Cov_results} shows the percentage of connected ground devices versus baseline approaches. In this figure, the exhaustive search method achieves the highest number of connected devices in both static (ES-S) and dynamic (ES-D) settings, with an average connection of about 96\% and 93\%, respectively. The exhaustive search method under the uneven and dynamic (ES-UD) setting had the least average connection of about 92\%. However, this method consumed the most energy in the the order of hundreds of KiloJoules as seen in Figure \ref{Fig:energy_results}. This high energy cost is due to the exploration of all possible combinations of the search space by the UAVs. The proposed DQLSI approach performs well in both static (DQLSI-S), dynamic (DQLSI-D) and uneven dynamic (DQLSI-UD) settings with average connections of about 93\%, 89\% and 88\%, respectively. Interestingly, the cluster-based Q-learning~\cite{Liu2019UAVqlearning} approach performs well with average connections of about 89\% in the static scenario, however, only about 33\% of the average connections were achieved in the dynamic setting. This poor performance was due to the inability of the UAVs to effectively locate the quickly-changing centroids in real-time.

Similar to the exhaustive search method, the iterative search method \cite{Mozaffari2017UAV} performs poorly in terms of energy consumption as seen in Figure \ref{Fig:energy_results}. Though it performs slightly better in the static case (IS-S) than in the dynamic case (IS-D) in the number of connected devices, with an average number of connections of 62\% and 48\%, respectively. From this figure, both learning-based approaches (DQLSI and CQL) perform well in minimizing the total energy consumed by UAV-BSs in the network, however, some outliers with poor energy utilization were encountered in their dynamic settings. The results in Figures \ref{Fig:Cov_results} and \ref{Fig:energy_results} show that DQLSI approach is robust in simultaneously improving the number of connections and minimizing the total energy consumed by UAV-BSs in both static and dynamic environments.

\vspace{-3mm}
\section{Conclusion}
This paper demonstrates that UAVs serving as UAV-BSs can learn via RL agents to cooperatively maximize the number of connected static and mobile ground devices while improving the energy utilization in the network. Unlike previous work, we take into account interference from neighbouring UAV cells and we do not assume global spatial knowledge of the location of ground devices. The performance of the proposed DQLSI approach was compared to state-of-the-art centralized approaches under realistic network conditions. Our proposed approach is robust in simultaneously improving the number of connections and minimizing the total energy consumed by UAV-BSs in both static and dynamic environments.

We pose pertinent questions that are open to further investigations. What impact will different distributions and mobility patterns have on the performance of our proposed approach? Considering the impact of communication cost, can agents learn to communicate with minimal information? Finally, can other cooperative methods offer faster convergence under dynamic network conditions? These are directions for future work.

\section*{Acknowledgment}
This work was supported, in part, by the Science Foundation Ireland (SFI) Grants No. 16/SP/3804 (Enable) and 13/RC/2077\_P2 (CONNECT Phase 2), as well as a research grant from SFI and the National Natural Science Foundation Of China (NSFC) under the SFI-NSFC Partnership Programme Grant Number 17/NSFC/5224.



%

\end{document}